\documentclass[preprint,showpacs,preprintnumbers,amsmath,amssymb]{revtex4}

\usepackage{graphicx}
\usepackage{epsfig}		
\usepackage{dcolumn}
\usepackage{bm}

\def\0{\mbox{\tiny $0$}}
\def\1{\mbox{\tiny $1$}}
\def\2{\mbox{\tiny $2$}}
\def\3{\mbox{\tiny $3$}}
\def\4{\mbox{\tiny $4$}}
\def\5{\mbox{\tiny $5$}}
\def\6{\mbox{\tiny $6$}}
\def\7{\mbox{\tiny $7$}}
\def\8{\mbox{\tiny $8$}}
\def\9{\mbox{\tiny $9$}}
\def\R{\mbox{\tiny $R$}}
\def\T{\mbox{\tiny $T$}}

\def\I{\mbox{\tiny $I$}}

\def\k{\mbox{\tiny $k$}}

\def\L{\mbox{\tiny $L$}}

\def\I{\mbox{\tiny $I$}}
\def\D{\mbox{\tiny $D$}}
\def\mi{\mbox{\tiny $-$}}
\def\pl{\mbox{\tiny $+$}}
\def\ppm{\mbox{\tiny $\pm$}}

\begin{document}

\title{Delay times for symmetrized and antisymmetrized quantum tunneling configurations}

\author{A. E. Bernardini}
\email{alexeb@ifi.unicamp.br}
\affiliation{Instituto de F\'{\i}sica Gleb Wataghin, UNICAMP,
PO Box 6165, 13083-970, Campinas, SP, Brasil.}
\altaffiliation[Also at] {~IST, Departamento de F\'{\i}sica, Av. Rovisco Pais, 1, 1049-001, Lisboa, Portugal.}

\date{\today}

\begin{abstract}
The transit times are obtained for a symmetrized  (two identical bosons) and an antisymmetrized (two identical fermions) quantum colliding configuration.
Considering two identical particles symmetrically impinging on a one-dimensional barrier, we demonstrate that the phase time and the dwell time give connected results where, however, the exact position of the scattered particles is explicitly determined by the phase time (group delay).
For the antisymmetrized wave function configuration, an unusual effect of {\em accelerated} transmission is clearly identified in a simultaneous tunneling of two identical fermions.
\end{abstract}

\date{\today}
\maketitle

\section{Introduction}

The problem of finding an accurate formulation for tunneling times and the correct interpretation of {\em superluminal} barrier tunneling opened up a fruitful discussion in the literature \cite{Hau89,But03,Win03A,Win03,Olk04} since pulses of light and microwaves appear to tunnel through a barrier at speeds grater than the ones of a reference pulse moving in the vacuum \cite{Nim92,Ste93,Hay01}.
If one measures the speed by the peak of the pulse, it looks faster than the incident wave packet.
The superluminal experiments that promoted these controversial discussions were performed by the mean of a lattice of layers of transparent and opaque materials arranged so that waves of some frequencies are reflected (through destructive interference) while other frequencies pass through the lattices giving rise to a Hartman-like filter effect \cite{Har62}.
The tunneling occurs when a wave impinges on a thin barrier of the opaque part of the multilayers and some small amount of the wave {\em leaks} through to the other side.
Only the leading edge of the incident wave packet survives the tunneling process without being severally attenuated to the point that it cannot be detected.
In all cases described by the non-relativistic (Schroedinger) dynamics \cite{Olk04}, the pulse (wave packet) that emerges from the tunneling process is greatly attenuated and front-loaded due to the {\em filter} effect. 

A possible explanation for such a phenomenon may be found in the analysis of the very rapid spreading of both initial and transmitted wave packets within the regime of large width momentum distribution.
Since the transmission probability ($T$) depends analytically on the momentum component $k$ ($T \equiv T(k)$), the initial (incident wave) momentum distribution can be completely distorted by the presence of the barrier of potential.
As there is no sharp beginning to a pulse, we cannot declare the instant of its arrival at a certain point since the distorted transmitted/reflected wave function destroys the stationary phase condition \cite{PBE} utilized for accurately determining the position of the transmitted/reflected components.

Using the procedure we call multiple peak decomposition \cite{Ber06,Ber04}, we demonstrate that superposing the amplitudes of the reflected and transmitted waves allows for the undistorted reconstruction of the scattered wave packets. We also notice that the proportion between the phase time and the time averaged stored energy can be used to explain the unusual effects of {\em accelerated} transmission \cite{Win03A,Win03}.
The transit times are obtained for a symmetrized  (bosons) and for an antisymmetrized (fermions) wave function which parameterizes the proposed colliding configuration.
In particular, we verify that the two identical fermion collision leads to an unusual effect of {\em accelerated} tunneling transmission which reactivates the possibility of superluminal tunneling for fermionic particles.

\section{Quantitative and qualitative analysis}

Let us consider a rectangular potential barrier $V(x)$, $V(x) = V_{\0}$ if $x \in \mbox{$\left[- L/2, \, L/2\right]$}$
and $V(x) = 0$ if $x \in\hspace{-0.3cm}\slash\hspace{0.1cm}\mbox{$\left[- L/2, \, L/2\right]$}$.
For what concerns the {\em standard} one-way direction wave packet tunneling, it is well-known \cite{Ber06} that the transmitted amplitude $T(k, L) = |T(k, L)|\exp{[i \Theta(k, L)]}$ is written in terms of
\small\begin{equation}
|T(k, L)| =
\left\{1+ \frac{w^4}{4 \, k^{\2} \, \rho^{\2}(k)}
\sinh^{\2}{\left[\rho(k)\, L \right]}\right\}^{-\frac{1}{2}},
\label{501}
\end{equation}\normalsize
and
\small\begin{equation}
\Theta(k, L) = \arctan{\left\{\frac{2\, k^{\2} - w^{\2}}
{2\,k \, \rho(k)}
\tanh{\left[2\,\rho(k) \, L \right]}\right\}},
\label{502}
\end{equation}\normalsize
for which we have made explicit the dependence on the barrier length $L$, and we have adopted $\rho(k) = \left(w^{\2} - k^{\2}\right)^{\frac{1}{2}}$ with $w = \left(2\, m \,V_{\0}\right)^{\frac{1}{2}}$ and $\hbar = 1$.
The above result is adopted for calculating the transit time $t_{T}$ of a transmitted wave packet when its peak emerges at $x = L/2$,
\small\begin{eqnarray}
t_{T} = \frac{m}{k_{\0}}\left.\frac{d\Theta(k, \alpha)}{dk}\right|_{_{k = k_{\0}}} =
\frac{2\,m \, L}{k_{\0} \,\alpha }
\left\{\frac{w^4\,\sinh{(\alpha)}\cosh{(\alpha)}
-\left(2\, k_{\0}^{\2} - w^{\2} \right)k_{\0}^{\2} \,\alpha }
{4\, k_{\0}^{\2} \,\left(w^{\2} - k_{\0}^{\2} \right)  +
w^4\,\sinh^{\2}{(\alpha)}}\right\}
\label{4}
\end{eqnarray}\normalsize
where $\alpha = w L \sqrt{(1 - k_{\0}^{\2}/w^{\2})}$ and $k_{\0}$ is the maximum of a generic symmetrical momentum distribution $g(k - k_{\0})$ building up the {\em incident} wave packet.
By following our previous analysis \cite{Ber06}, it is well-established that, due to the {\em filter effect}, the amplitude of the transmitted wave is essentially composed by the plane wave components of the front tail of the {\em incoming} wave packet which reaches the first barrier interface before the peak arrival \cite{Lan89}.
We have shown that the {\em cut off} of the momentum distribution at $k \simeq (1 - \delta) w$ increases the amplitude of the tail of the incident wave so that it contributes so relevantly as the peak of the incident wave to the final composition of the transmitted wave.
Independently, due to the novel asymmetric character of the transmitted amplitude $g(k-k_{\0}) |T(k, L)|$, an ambiguity in the definition of the {\em arrival}/{\em transmitted} time is created \cite{Ber06}.
In the framework of the multiple peak decomposition \cite{Ber04}, we have suggested a suitable way for comprehending the conservation of probabilities
where the asymmetric aspects previously discussed \cite{Ber06} could be totally eliminated.
By considering the same rectangular barrier $V(x)$, we solve the Schroedinger equation for a plane wave component of momentum $k$ for two identical wave packets symmetrically separated from the origin $x = 0$.
By assuming that $\phi^{\L(\R)}(k,x)$ are Schroedinger equation stationary wave solutions,
when the wave packet peaks simultaneously reach the barrier (at the mathematically convenient time $t = - (m L) /(2 k_{\0})$) we can write
\small\begin{equation}
\phi^{\L(\R)}(k,x)=
\left\{\begin{array}{l l l l}
\phi^{\L(\R)}_{\1}(k,x) &=&
\exp{\left[ \pm i \,k \,x\right]} + R^{\L(\R)}(k,L)\exp{\left[ \mp i \,k \,x\right]}&~~~~x < - L/2\, (x > L/2),\nonumber\\
\phi^{\L(\R)}_{\2}(k,x) &=& \gamma^{\L(\R)}(k)\exp{\left[ \mp\rho  \,x\right]} + \beta^{\L(\R)}(k)\exp{\left[ \pm\rho  \,x\right]}&~~~~- L/2 < x < L/2,\nonumber\\
\phi^{\L(\R)}_{\3}(k,x) &=& T^{\L(\R)}(k,L)\exp{ \left[\pm i \,k \,x\right]}&~~~~x > L/2 \, (x < - L/2) .
\end{array}\right.
\label{510}
\end{equation}\normalsize
where the upper(lower) sign is related to the index $L$($R$) corresponding to the incidence on the left(right)-hand side of the barrier.
By assuming the conditions for the continuity of $\phi^{\L,\R}$ and their derivatives at $x = - L/2$ and $x = L/2$, after some mathematical manipulations, we can easily obtain for the reflection amplitude $R^{\L,\R}(k,L)$,
\small\begin{equation}
R^{\L,\R}(k,L) = \exp{\left[ - i \,k \,L \right]} \left\{\frac{\exp{\left[ i \, \theta(k)\right]} \left[1 - \exp{\left[ 2\,\rho(k) \,L\right]}\right]}{1 - \exp{\left[ 2\,\rho(k) \,L\right]}\exp{\left[ i\, 2\,\theta(k)\right]}}\right\},
\label{511}
\end{equation}\normalsize
and for the transmission amplitude $T^{\L,\R}(k,L)$,
\small\begin{equation}
T^{\L,\R}(k,L) = \exp{\left[ - i \,k \,L \right]} \left\{\frac{\exp{\left[\rho(k) \,L\right]}\left[1- \exp{\left[ 2\, i\, \theta(k)\right]}\right]}{1 - \exp{\left[ 2\,\rho(k) \,L\right]}\exp{\left[ i\, 2\,\theta(k)\right]}}\right\},
\label{512}
\end{equation}\normalsize
where
\small\begin{equation}
\theta(k) = \frac{ 2\, k \, \rho(k)}{2k^{\2} - w^{\2}}.
\label{512B}
\end{equation}\normalsize
$R^{\L(\R)}(k,L)$ and $T^{\R(\L)}(k,L)$ are intersecting each other.

Since the above colliding configuration is spatially symmetric, the symmetry operation corresponding to the $1 \leftrightarrow 2$ particle exchange can be parameterized by the position coordinate transformation $x \rightarrow -x$.
At the same time, it is easy to observe that
\small\begin{equation}
\phi^{\L(\R)}(k,x) = \phi^{\L(\R)}_{\1 \pl \2 \pl \3}(k,x) = \phi^{\R(\L)}_{\1 \pl \2 \pl \3}(k,-x) = \phi^{\R(\L)}(k,-x)
\label{512C}
\end{equation}\normalsize
where the $L \leftrightarrow R$ interchange is explicit.
Consequently, in case of analyzing the collision of two identical bosons, we have to consider a symmetrized superposition of the $L$ and $R$ wave functions,
\small\begin{equation}
\phi_{\pl}(k,x) = \phi^{\L}(k,x) + \phi^{\R}(k,x) = \phi^{\R}(k,-x) + \phi^{\L}(k,-x) = \phi_{\pl}(k,-x).
\label{512D}
\end{equation}\normalsize
Analogously, in case of analyzing the collision of two identical fermions (just taking into account the spatial part of the wave function),
we have to consider an antisymmetrized superposition of the $L$ and $R$ wave functions,
\small\begin{equation}
\phi_{\mi}(k,x) = \phi^{\L}(k,x) - \phi^{\R}(k,x) = \phi^{\R}(k,-x) - \phi^{\L}(k,-x) = -\phi_{\mi}(k,-x).
\label{512E}
\end{equation}\normalsize
Thus the amplitude of the re-composed transmitted plus reflected waves would be given by
$R^{\L,\R}(k,L) + T^{\R,\L}(k,L)$ for the symmetrized wave function $\phi_{\pl}$ and by
$R^{\L,\R}(k,L) - T^{\R,\L}(k,L)$ for the antisymmetrized wave function $\phi_{\mi}$.
Resorting to the multiple peak decomposition \cite{Ber04} applied to such a pictorial symmetrical tunneling configuration,
we can superpose the amplitudes of the intersecting probability distributions before taking their squared modulus in order to obtain
\small\begin{eqnarray}
R^{\L,\R}(k,L) \pm T^{\R,\L}(k,L)
&=&
 \exp{\left\{ - i [k \,L - \varphi_{\pm}(k,L)]\right\}}
\label{513}
\end{eqnarray}\normalsize
with
\small\begin{equation}
\varphi_{\pm}(k,L) = - \arctan{\left\{\frac{2\,k\,\rho(k) \, \sinh{[\rho(k)\,L]}}{\left(k^{\2}-\rho^{\2}(k)\right)\cosh{[\rho(k)\,L]} \pm w^{\2}}\right\}},
\label{514}
\end{equation}\normalsize\normalsize
where the {\em plus} sign is related to the results obtained for the a symmetrized superposition and the {\em minus} sign is related to the antisymmetrized superposition.
From Eq.~(\ref{513}) it is immediate that that $|R^{\L,\R}(k,L)\pm T^{\R,\L}(k,L)| = 1$; then in both odd and even wave function symmetrization cases, the original undistorted distribution is recovered.
The previously pointed out incongruities which cause the distortion of the momentum distribution $g(k - k_{\0})$ are completely eliminated
and we recover the fundamental condition for the applicability of the SPM for accurately determining the position of the peak of the reconstructed wave packet composed by reflected and transmitted superposing components.
The phase time interpretation can be, in this case, correctly quantified in terms of the analysis of the novel phase $\varphi_{\pm}(k, L)$ since
the novel scattering amplitudes $g(k - k_{\0})|R^{\L,\R}\pm T^{\R,\L}| \simeq g(k - k_{\0})$ maintains its previous symmetrical character.
The transmitted and reflected interfering amplitudes results in a unimodular function which just modifies the {\em envelop} function $g(k - k_{\0})$ by an additional phase.
and the scattering phase time results in
\small\begin{equation}
t^{(\alpha)}_{T, \varphi_{\ppm}} =\frac{m }{k}\frac{d\varphi(k, \alpha)}{dk} =
\frac{2\,m\, L}{k\,\alpha}
\frac{w^{\2}\sinh{(\alpha)} \pm \alpha\,k^{\2}}{2\,k^{\2} - w^{\2} \pm w^{\2}\cosh{(\alpha)}}
\label{515}
\end{equation}\normalsize
where $k \rightarrow k_{\0}$, with $\alpha$ previously defined.
The {\em old} phase $\Theta(k, L)$ (Eq.~\ref{502}) appears when we treat separately the momentum amplitudes
$T(k, L)$ and $R(k, L)$, which destroys the symmetry of the initial momentum distribution $g(k - k_{\0})$ by the presence of the
multiplicative term $T(k, L)$ or $R(k, L)$, and the novel phase $\varphi(k, L)$ appears only when we sum the tunnneling/scattering amplitudes so that the symmetrical character of the initial momentum distribution is recovered (due to the result of Eq.~(\ref{513})).

To illustrate the difference between the {\em standard} tunneling phase time $t^{(\alpha)}_{T}$ and the {\em symmetrical} scattering phase time $t^{(\alpha)}_{T, \varphi}$ we introduce the parameter $n = k^{\2}/w^{\2}$ and we define the {\em classical} traversal time $\tau_{\k} = (m L) /k$.
\begin{figure}[th]
\vspace{-0.6 cm}
\centerline{\psfig{file=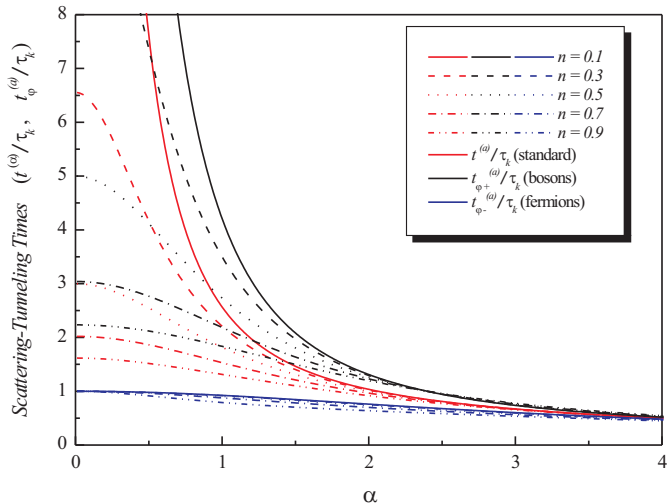,width=10cm}}
\vspace{-1 cm}
\caption{Normalized phase times for a {\em symmetrical} tunneling configuration
of the symmetrized wave function representing the collision of two identical bosons (black lines)
and the antisymmetrized wave function representing the collision of two identical fermions (red lines).
The results are plotted in comparison with the {\em standard} one-way direction tunneling phase times (blue lines).
These times can be understood as transit times in the units of the {\em classical} traversal time $\tau_{\k} = (m L) /k$.
All the above phase time definitions present the same asymptotic behavior.
\label{fig3A}}
\vspace{-0.4 cm}
\end{figure}
In this case, we can obtain the normalized phase times
\small\begin{equation}
t^{(\alpha)}_{T}
=
\frac{2  \tau_{\k}}{\alpha}\left\{
\frac{\cosh{(\alpha)}\sinh{(\alpha)} - \alpha\,n\left(2 n - 1\right)}{\left[4 n \left(1 - n\right)+\sinh^{\2}{(\alpha)}\right]}
\right\}
\label{517A}
\end{equation}\normalsize
and
\small\begin{equation}
t^{(\alpha)}_{T, \varphi_{\pm}}
=
\frac{2  \tau_{\k}}{\alpha}\left\{\frac{n\, \alpha \pm \sinh{(\alpha)}}{2n - 1 \pm\cosh{(\alpha)}}
\right\}.\label{517}
\end{equation}\normalsize

\begin{figure}[th]
\vspace{-0.6 cm}
\centerline{\psfig{file=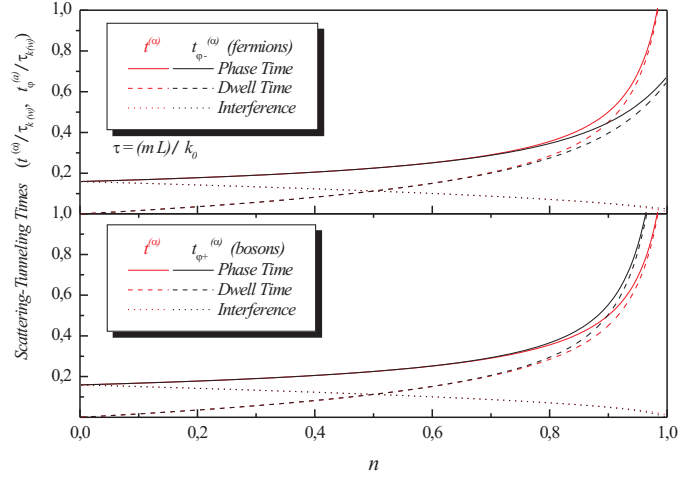,width= 10cm}}
\vspace{-1 cm}
\caption{Exact phase time (solid line), self-interference delay (dotted line), and the dwell time (dashed line)
as a function of the normalized energy $n = k^{\2}/w^{\2} \propto E_{\0}/V_{\0}$ for the two identical particles (black line) and the standard one-way direction (red line)
collision with a rectangular potential barrier.
For the colliding two identical fermions (first plot) we have assumed that the wave function is totally antisymmetrized.
For the colliding two identical bosons (second plot) we have assumed that the wave function is totally symmetrized.
These times are normalized by the {\em classical} traversal time $\tau_{\k} = (m L) /k$, and here we have adopted $w L = 4\pi$ for $\alpha = w L\sqrt{1-n}$.
\label{fig2}}
\vspace{-0.4 cm}
\end{figure}

At this point, one could say metaphorically that two bosonic particles represented by the symmetrized incident wave function spend a time equal to $ t_{T, \varphi_{\pl}}$ inside the barrier before retracing its steps or tunneling and that two fermionic particles represented by antisymmetrized incited wave function spend a time equal to $ t_{T, \varphi_{\mi}}$.
The physical realization of such a metaphor relies on the definition of the dwell time for the same colliding configuration which we have proposed.
The dwell time is a measure of the time spent by a particle in the barrier region regardless of whether it is ultimately transmitted or reflected \cite{But83},
\small\begin{equation}
t_{D,\ppm}
=\frac{m}{k} \int_{\mi \L/\2}^{{\pl \L/\2}}\mbox{d}x{|\phi_{\pm,\2}(k,x)|^{\2}}
\label{530}
\end{equation}\normalsize
where  $j_{in}$ is the flux of incident particles and $\phi_{\2}(k,x)$ is the stationary state wave function depending on the colliding configuration that we are considering (symmetrical or standard).
To derive the relation between the dwell time and the phase time, we reproduce the variational theorem which yields the sensitivity
of the wave function to variations in energy.
After some elementary manipulations of the Schroedinger equation \cite{Smi60}, we can write
\small\begin{equation}
\phi^{\dagger}\phi = \frac{1}{2m}\frac{\partial}{\partial x}\left(\frac{\partial \phi}{\partial E}\frac{\partial \phi^{\dagger}}{\partial x} - \phi^{\dagger}\frac{\partial^{\2}\phi}{\partial E\partial x}\right).
\label{531}
\end{equation}\normalsize
Upon integration over the length of the barrier we find
\small\begin{equation}
2 m \int_{\mi \L/\2}^{{\pl \L/\2}}\mbox{d}x{|\phi_{\2,\ppm}(k,x)|^{\2}} = \left.\left(\frac{\partial \phi}{\partial E}\frac{\partial \phi^{\dagger}}{\partial x} - \phi^{\dagger}\frac{\partial^{\2}\phi}{\partial E\partial x}\right)\right|_{\mi\L/\2}^{\pl\L/\2}.
\label{532}
\end{equation}\normalsize
At the barrier limits ($x = \pm L/2$), for the symmetrical configuration that we have proposed, we can use the superposition of the scattered waves to explicitly calculate 
\small\begin{eqnarray}
\left.\phi_{\ppm}(k,x)\right|_{\mi\L/\2(\pl\L/\2)} &=& \frac{\phi^{\L(\R)}_{\1}(k,x) \pm \phi^{\R(\L)}_{\3}(k,x)}{\sqrt{2}}
\nonumber\\
&=&
  \exp{\left[ \pm i \,k \,x\right]} + \exp{\left[ \mp i \,k \,x + i \left(\varphi_{\ppm}(k,L)- k L\right)\right]}
\label{533}
\end{eqnarray}\normalsize
By evaluating the right-hand side of the Eq.~(\ref{533}), we obtain
\small\begin{equation}
\frac{\partial k}{\partial E} \frac{d\varphi_{\ppm}}{dk} = \frac{m}{k} \int_{\mi \L/\2}^{{\pl \L/\2}}\mbox{d}x{|\phi_{\2,\ppm}(k,x)|^{\2}}
- \frac{Im[\exp{(i \varphi_{\ppm})}]}{k} \frac{\partial k}{\partial E}.
\label{534}
\end{equation}\normalsize
The first contribution to the right-hand side of Eq.~(\ref{534}) represents the phase time, the second one leads to the explicit computation of the dwell time.
By imposing the continuity conditions of the Schroedinger equation solutions,
in the barrier region we obtain a stationary wave symmetrical or antisymmetrical in $x$,
\small\begin{eqnarray}
\phi_{\2,\ppm}(k,x) &=& \frac{\phi^{\L}_{\2}(k,x) \pm \phi^{\R}_{\2}(k,x)}{\sqrt{2}}~~~~~~(\gamma\equiv\gamma^{\L,\R}\, \beta\equiv\beta^{\L,\R})
\nonumber\\
&=&
\sqrt{2}(\beta + \gamma)\frac{\exp{[\rho(k)\,x]} \pm \exp{[\rho(k)\,x]}}{2},
\label{534B}
\end{eqnarray}\normalsize
which, from Eq.~(\ref{530}), leads to
\small\begin{equation}
t^{(\alpha)}_{D, \varphi_{\ppm}} =
\frac{2\, \tau_{\k}\, n}{\alpha}
\frac{\alpha\pm\sinh{(\alpha)}}{2n - 1 \pm \cosh{(\alpha)}}
\label{535}
\end{equation}\normalsize
The self-interference term which comes from the momentary overlap between the incident and the reflected waves in front of the barrier is given by\footnote{We have printed the phase index $\varphi$ for all the results related to the symmetrical colliding configuration.}
\small\begin{eqnarray}
t^{(\alpha)}_{\I, \varphi_{\ppm}} &=& - \frac{Im[\exp{(i \varphi_{\ppm})}]}{k} \frac{\partial k}{\partial E}
=  \frac{m \, \sin{(\varphi_{\ppm})}}{k^{\2}} = \pm\frac{2 \tau_{\k}}{\alpha} \frac{(1-n)\sinh{(\alpha)}}{2n - 1 \pm \cosh{(\alpha)}}.
\label{536}
\end{eqnarray}\normalsize
It is interesting to observe that the result for the self-interference delay (\ref{536}) $t^{(\alpha)}_{\I, \varphi_{\ppm}}$ in the above equation also depends on the parity of the wave function.
The dwell time is obtained from a simple subtraction of the quote self-interference delay
$t_{\I, \varphi_{\ppm}}$ from the phase time that describes the exact position of the peak of the scattered wave packets, i. e. $t_{\T,\varphi_{\ppm}} = t_{\D, \varphi_{\ppm}}+ t_{\I, \varphi_{\ppm}}$
as we can notice in the Fig.~\ref{fig2}.
Adopting the {\em classical} traversal time $\tau_{\k} = (m L) /k$ for normalizing the results displayed in Fig.~\ref{fig2} allows us to point out a crucial aspect regarding two identical fermions collision, that is the possibility of an unusual accelerated tunneling transmission.
In fact, when each separated transmission coefficient $T^{\L,\R}$ prevails over the each reflection coefficient $R^{\L,\R}$, i. e. $|T|^{\2} > |R|^{\2}$, we have $|T|^{\2} > 1/2$.
For satisfying such a requirement the Eq.~(\ref{501}) gives $(w L)/(2\sqrt{n}) \leq w L \sinh{(\alpha)}) /(2\sqrt{n}) < 1 $.
For two identical bosonic particles, the possibility of accelerated tunneling transitions with respect to the traversal
{\em classical} course is quantified by $t_{T, \varphi_{\pl}}^{(\alpha)}< \tau_{\k}$
It occurs only when $\alpha/2 \geq (\alpha/2) \tanh{(\alpha/2)} > 1$.
Since $\alpha = w L\sqrt{1-n}$, the intersection of the ``weak version'' of both of the above constraints, $(w L)/(2\sqrt{n}) < 1$ and $\alpha/2 > 1$,
leads to $n > 2$, which definitely does not correspond to an effective tunneling configuration.
In the region where the one-way direction transmission coefficient dominates on the reflection coefficient, bosons should tunnel with a retarded velocity with respect to the classical velocity since we have, in this case, $t^{(\alpha)}_{T, \varphi}> \tau_{\k}$.
It does not correspond to the theoretical results for two fermionic particles, for which $t_{\T, \varphi_{\mi}}^{(\alpha)}$ is always smaller than $\tau_{\k}$, because it is not possible to establish a link between the relation $t_{\T, \varphi_{\mi}}^{(\alpha)} < \tau_{\k}$ and the coefficients $R$ and $T$.
Consequently, the possibility of accelerated tunneling and the verifiability of the Hartman effect for the two identical fermion tunneling configuration is, in fact, concrete.

\section{Conclusion and outlook}

We have demonstrated that the phase time and the dwell time give connected results in spite of the exact position of the scattered particles being explicitly given by the phase time (group delay).
For the antisymmetrized (two identical fermions) wave function configuration, an unusual effect of {\em accelerated} tunneling effect have been clearly identified in simultaneous two identical fermion tunneling.
Even with the introduced modifications, our results partially corroborate the analysis of Refs. \cite{Win03A,Win03} that gives an answer to the paradox of the Hartman interpretation \cite{Har62}.
In particular, we provide a way of comprehending the conservation of probabilities \cite{Ber04, Ber06} for a very particular tunneling configuration where the asymmetry and the distortion aspects presented in the standard case were all eliminated.
Otherwise, one should keep in mind that {\em accelerated} tunneling transmission and, generically, the Hartman effect, even in its more sophisticate consequences, appear to have been experimentally verified, in particular, for opaque barriers and nonresonant tunneling \cite{Zai05}, and, under severe analytical restrictions, reproduced also by numerical simulations and constrained theoretical analysis \cite{Pet03}.
Due to the importance and the eventual correlated experimental verifiability of such tunneling time measurements in nanostructures like single- and bi-layer graphene, where antisymmetrized wave functions can be mimicked by electron-hole pairs, the construction proposed in this letter could certainly be extended to the pseudo-relativistic dynamics of electrons in graphene.
In this sense, we suggest that the results here obtained require further attention by experimenters on non-symmetrical quantum tunneling constructions.
{\bf Acknowledgments}
We would like to thank FAPESP (PD 04/13770-0) for the financial support.

\end{document}